\documentclass[11pt,a4paper]{article}
\usepackage{jheppub}
\usepackage[T1]{fontenc}
\usepackage{amssymb}
\usepackage{mathtools}
\usepackage{stackengine}
\usepackage{scalerel}

\usepackage{hyperref}
\usepackage{epsfig}
\usepackage{amsmath,latexsym,amssymb}
\usepackage{graphicx}
\usepackage{braket}
\usepackage{pdfsync}

\usepackage{framed}
\usepackage{mathtools}

\usepackage{jheppub}
\usepackage[T1]{fontenc}
\usepackage{amssymb}
\usepackage{mathtools}
\usepackage{amsmath}
\usepackage{bm}
\usepackage{stackengine}

\usepackage{scalerel}
\usepackage{eufrak}
\usepackage{framed}

\usepackage{mathtools}

\usepackage{pstricks}

\usepackage{amsmath}
\usepackage{bbold}
\usepackage{bm}
\usepackage{latexsym}
\usepackage{braket}
\usepackage{slashed}
\usepackage{graphicx,booktabs,multirow}
\usepackage{eufrak}
\numberwithin{equation}{section}

\usepackage{color}
\usepackage{pstricks}
\usepackage{amssymb}

\makeatletter
\newcommand{\doublewidetilde}[1]{{%
  \mathpalette\double@widetilde{#1}%
}}
\newcommand{\double@widetilde}[2]{%
  \sbox\z@{$\m@th#1\widetilde{#2}$}%
  \ht\z@=.9\ht\z@
  \widetilde{\box\z@}%
}
\makeatother

\usepackage{hyperref}
\usepackage{slashed}
\usepackage{epsfig}
\usepackage{amsmath,latexsym,amssymb}
\usepackage{graphicx}
\usepackage[latin1]{inputenc}
\usepackage{braket}
\usepackage{pdfsync}

\usepackage{framed}

\usepackage{mathtools}

\def\be{\begin{equation}}
\def\ee{\end{equation}}
\def\ba{\begin{eqnarray}}
\def\ea{\end{eqnarray}}

\def\ket#1{\mathinner{|{#1}\rangle}}

\def\braket#1{\mathinner{\langle{#1}\rangle}}

\renewcommand{\braket}[2]{\left< #1 \vphantom{#2} \right|
 \left. #2 \vphantom{#1} \right>} 
\newcommand{\matrixel}[3]{\left< #1 \vphantom{#2#3} \right|
 #2 \left| #3 \vphantom{#1#2} \right>} 

\newcommand{\comment}[1]{}

\newcommand{\eea}{\end{eqnarray}}

\author{
Tomasz R.\ Taylor${}^{1,2}$,\, Bin Zhu${}^{3}$\\[0.5cm]
 $^1${\it Department of Physics,
  Northeastern University, Boston, MA 02115, USA}\\
  $^2${\it Faculty of Physics, University of Warsaw, Pasteura 5, 02-093 Warsaw, Poland}\\
$^3${\it School of Mathematics and Maxwell Institute for Mathematical Sciences, University of Edinburgh,
EH9 3FD, UK }\\[0.2cm]
}

\emailAdd{taylor@neu.edu}
\emailAdd{bzhu@exseed.ed.ac.uk}

\title{\boldmath {Scattering of Quantum Particles in  Global de Sitter Spacetime~I:~ The Formalism} \unboldmath}

\abstract{We develop a formalism for computing the scattering amplitudes in maximally symmetric de Sitter spacetime with compact spatial dimensions. We describe quantum states by using the representation theory of de Sitter symmetry group and link the Hilbert space to ``inertial'' geodesic observers. The positive and negative ``energy'' wavefunctions are uniquely determined by the requirement that  in  observer's neighborhood, short wavelengths propagate as plane waves with positive and negative frequencies, respectively; they define  a unique ``Euclidean'' (a.k.a.\ Bunch-Davies) de Sitter invariant vacuum, common to all inertial observers.
By following the same steps as in Minkowski spacetime, we show that the scattering amplitudes are given by a generalized Dyson's formula.
Compared to the flat case, they describe the scattering of wavepackets with the frequency spectrum determined by geometry.
The frequency spread shrinks as  the masses and/or momenta become larger than the curvature scale. Asymptotically, de Sitter amplitudes agree with the amplitudes evaluated in Minkowski spacetime.}

\gdef\@fpheader{}
\makeatother

\begin{document}
\maketitle
\section{Introduction}

Elementary particle physics strives to explore nature at shortest possible distances. The Large Hardron Collider (LHC) probes distances as short as $\hbar c/(10\rm \,TeV)  \approx 10^{-20}$m. This is the realm of quantum physics, governed by the Planck constant, at microscopic distances 46 orders of magnitude shorter than the Hubble length!

Elementary particles are propagating in spacetime which is curved due to the enormous energy stored in matter, dark energy and radiation.
The curvature changes the kinematics. It affects the trajectories of incoming and outgoing particles. Since the distances probed in  the collisions are many orders of magnitude shorter than the cosmological distances, one usually assumes that the effects of curvature are negligible in accelerator and laboratory settings. They were were never analyzed, however, in a rigorous manner. The main challenge is the computation of the scattering amplitudes, thus generalizing the S-matrix from Minkowski to curved spacetime.

There have been many years of controversy regarding quantum fields in de Sitter spacetime. Already at the very outset, in the simplest case of scalar fields, one encounters ambiguities in the definition of the vacuum state \cite{Mottola:1984ar,Allen:1985ux}.
The difficulties pile up in each step towards constructing a self-consistent S-matrix-like object, so it is not surprising that, as phrased by Marolf, Morrison and Srednicki in Ref.\cite{Marolf:2012kh}, ``Arguments continue over the interpretation of calculations in both simple self-interacting scalar theories on a fixed de Sitter background (...) and also in the more complicated case of gravitational theories...'' Act{\nolinebreak}ually, Ref.\cite{Marolf:2012kh} contains a comprehensive list of (still) open problems, together with some representative references. We do not repeat them here because  we can only make it worse by adding our two cents. Instead, we start from the scratch and focus on the notion what de Sitter observers would see as their proper time elapses. The observer plays a central role if the construction proposed in this work.

In Minkowski spacetime, the construction of S-matrix relies heavily on Poincar\'e symmetry, with a preferred class of coordinate systems of ``inertial'' observers moving along straight world-lines, who can transform the kinematic variables and amplitudes from one reference frame to another by using Lorentz transformations. In $d$ dimensions, maximally symmetric de Sitter spacetime deS$_d$ enjoys $SO(1,d)$  symmetry which has the same dimension ${d(d+1)}/{2}$ as the Poincar\'e group generated by ${(d-1)d}/{2}$ Lorentz transformations and $d$ translations in $d$ flat spacetime dimensions. This symmetry relates de Sitter geodesics in the same way as Poincar\'e symmetry relates inertial observers. Hence it is not surprising that, as elaborated in this work, we can construct  de Sitter scattering amplitudes by repeating, step by step, the standard procedures of S-matrix formalism in flat spacetime.
 In our construction, Hilbert space will be linked to the observers ``living'' on timelike geodesics and  the scattering amplitudes
will be measured  in their reference frames.

This paper is organized as follows. In Section 2,  we describe general features
de Sitter spacetime, with $SO(1,d)$ symmetry inherited from $(d{+}1)$-dimensional embedding space. We describe the geodesics of observers who measure the scattering amplitudes. In Section 3, we discuss the symmetry algebra of physical observables and their flat limits obtained by the In\"on\"u-Wigner contraction. We construct the Hamiltonian operator generating proper time translation along timelike geodesic lines.
In Section 4, we discuss scalar field theory in de Sitter spacetime. In order to make it as explicit as possible, we specify to two dimensions and choose the mass above the curvature scale. By considering the short wavelength limit of Klein-Gordon solutions, we identify the positive and negative frequency wavefunctions. This leads to a unique split of the scalar field into the positive and negative frequency modes and consequently, to the identification of the creation and annihilation operators. The vacuum, annihilated by the annihilation operators, is unique and common to all observers (invariant under de Sitter symmetry). In Section 5, we  perform the spectral decomposition of scalar wavefunctions. We discuss the frequency distribution in the limits of large mass and momentum, when the particles probe de Sitter spacetime at short invariant intervals. These distributions peak at the frequencies expected for particles propagating in flat spacetime.
In Section 6, we identify the symmetry generators as Noether charges of the currents constructed from the stress-energy tensor and Killing vectors. It allows expressing the Hamiltonian of interacting theory in terms of the stress-energy tensor.
In Section 7, we use this Hamiltonian to derive a general expression for the scattering amplitudes. To that end, we follow the same steps as in Minkowski spacetime and derive a curved space generalization of the familiar Dyson's formula.
Compared to the flat case, the amplitudes describe the scattering of wavepackets with the frequency spectrum determined by geometry.  In the limit of  masses and/or momenta far above the curvature scale, de Sitter scattering amplitudes agree with the amplitudes evaluated in Minkowski spacetime. We illustrate it on a simple example in Section 8. In Section 9, we summarize our results and compare with the existing literature.

Although most of our arguments apply to any number of dimensions, we often use two dimensions as an example, to make the reasoning as transparent as possible. The computations involve a plethora of special functions and their asymptotic expansions. Most of formulas necessary to reproduce our calculations can be found in NIST Digital Library of Mathematical Functions https://dlmf.nist.gov/ - only in rare cases, we refer to the original literature. We also included a short appendix with the definitions and some properties of associated Legendre functions which we often use in our computations.

\section{Observers in maximally symmetric de Sitter space}
de Sitter spacetime can be constructed in any number of dimensions $d$.  It may be realized as a hypersurface described by the equation
\be -X_0^2+X^2_1+\dots +X^2_d=\ell^2\ee
in flat $(d+1)$-dimensional Minkowski space \cite{Spradlin:2001pw}, see Fig.\ 1.
The $SO(1,d)$ de Sitter isometry group follows from Lorentz symmetry of the constraint.
The parameter $\ell$ with units of length is called de Sitter radius.
It is related to the  curvature scalar and the related cosmological constant in the following way:
\be R={d(d-1)\over \ell^2}\ ,\qquad
\Lambda={(d-1)(d-2)\over 2\ell^2} \, .\ee
Note that $\Lambda=0$ in $d=2$ while $R=2/\ell^2\neq 0$, which is due to the vanishing Einstein tensor in two dimensions. Actually deS$_2$
\cite{Sun:2021thf}, which is easy to visualize as a two-dimensional hyperboloid, captures the most important features of four-dimensional de Sitter spacetime and will serve us as a ``proof of concept.''
\begin{figure}\centering\includegraphics[scale=0.15,page=1]{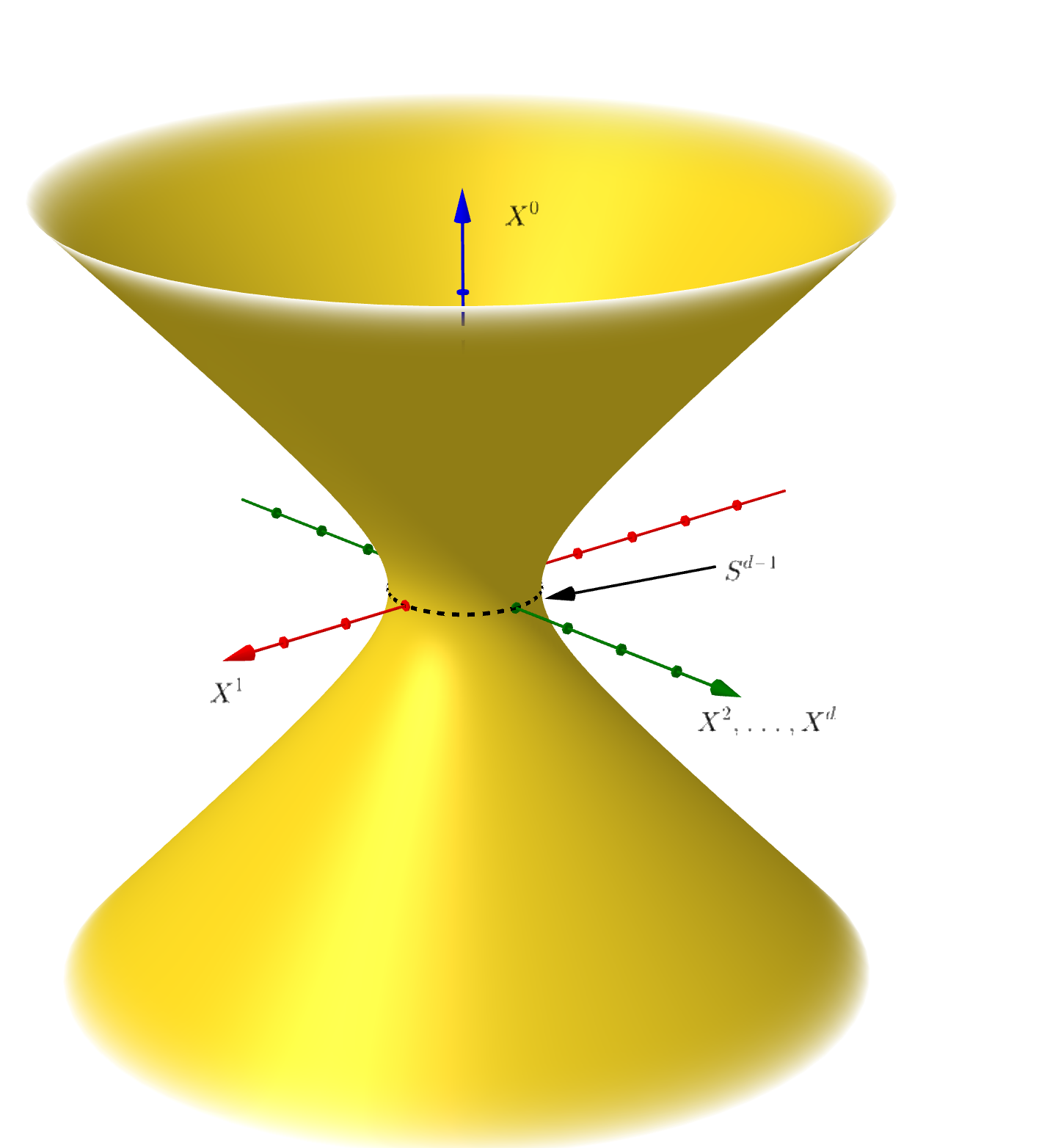}\caption{deS$_d$ in embedding $(d{+}1)$-dimensional  Minkowski space.}\end{figure}
\begin{figure}\centering\includegraphics[scale=0.15,page=1]{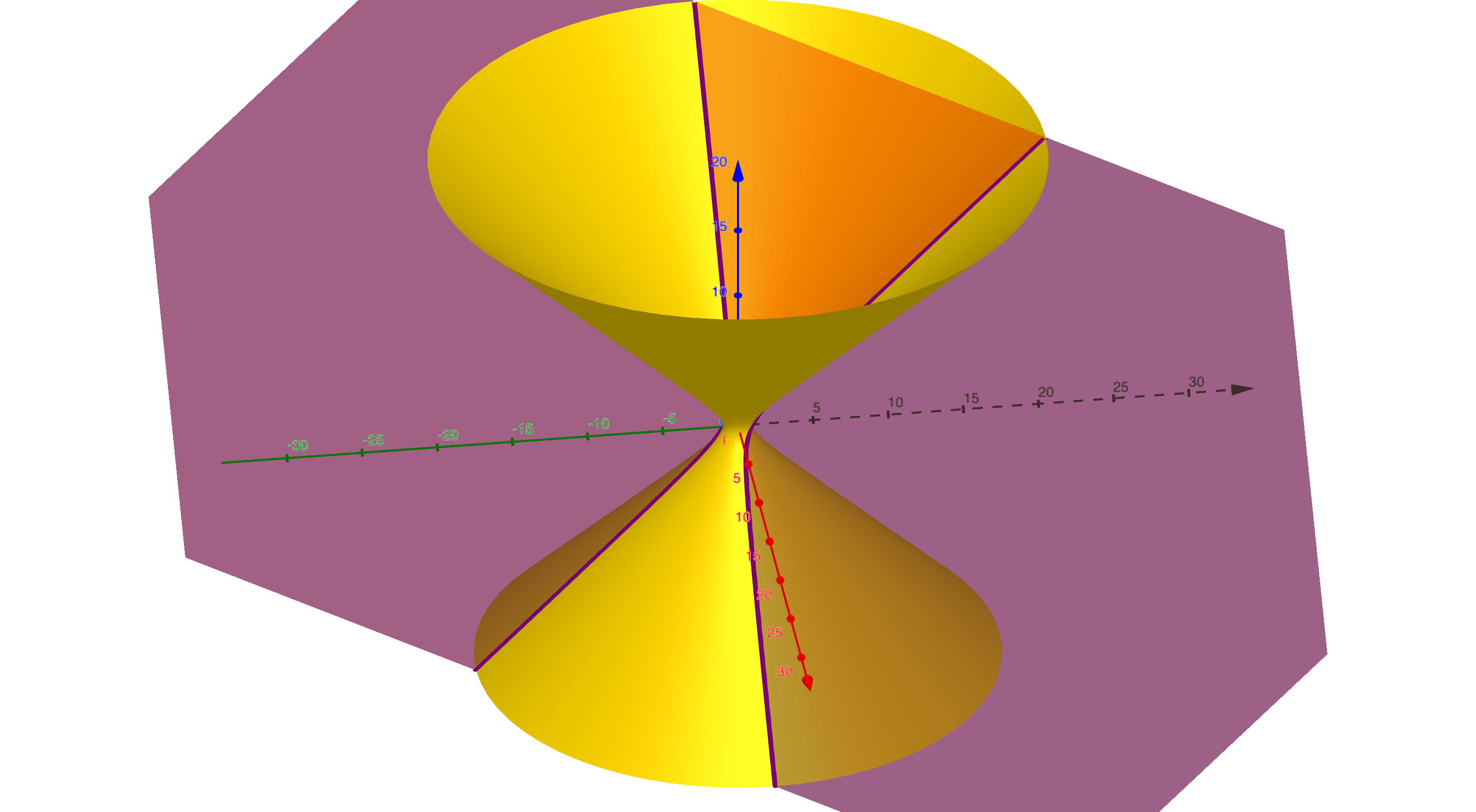}\vskip 0 cm\caption{Geodesic lines carved out by a plane.}\vskip -0 cm\end{figure}

The $SO(1,d)$ de Sitter isometry group is generated by the (hermitean) Lorentz gener{\nolinebreak}ators $L_{AB}=-L_{BA} ~ (A,B=0,1,\dots d)$ of the embedding space, which obey the algebra
\be [L_{AB},L_{CD}]=i\, (\eta_{BC}L_{AD} -\eta_{AC}L_{BD} +\eta_{AD}L_{BC}-
\eta_{BD}L_{AC}) \label{alg},\ee
where $\eta_{AB}=\rm{diag}(-1,1,\dots,1)$. 

The observers and massive particles follow timelike geodesics. These be constructed  as  the intersections between deS$_d$ and two-dimensional planes containing the origin of the embedding space and having $d-1$ independent spacelike normal vectors \cite{Cacciatori:2007in}, see Fig.\ 2. Each plane intersects the future lightcone along two of its generatrices. Any two future directed null vectors $v$ and $w$ lying on
such generatrices (and normalized as $2v\cdot w =-1$) can be used to parameterize the corresponding geodesic in
terms of the proper time $\tau$:
\be X(\tau)=\ell(v \,e^{c\tau/\ell}-w \,e^{-c\tau/\ell})\  .\label{ged}\ee
Actually, the same plane carves out another ``mirror'' geodesics,
$X'(\tau)= -X(\tau)$.

The  geodesics are characterized (and uniquely determined) by $d(d+1)/2$  conserved Noether charges $K_{AB}=-K_{BA}$  associated to the Lorentz generators $L_{AB}$.
{}For a motion parameterized by
 the vectors $v$ and $w$  as in in Eq.(\ref{ged}), they are given by \cite{Cacciatori:2007in}
\be K_{AB}=v_Aw_B-w_Av_B\ .\label{kab}\ee 
Similarly to Minkowski spacetime, where the observers are related by Poincar\'e transformations, de Sitter observers are related by $SO(1,d)$ transformations. In particular, each geodesic motion can be transformed into the motion on a ``vertical'' geodesics in the (01) plane, with $X_A(\tau)=0$ for $A>1$, and the observer's $X_0(\tau=0)=0$, $X_1(\tau=0)=\ell$.

deS$_d$ is topologically equivalent to $\mathbb{R}\times S^{d-1}$. In the ``decompactification'' limit of $\ell\to\infty$, deS$_d$ becomes flat.  The compact dimensions, however, as well as the lower rank of the symmetry algebra as compared to the flat case, leave a mark on the Hilbert space.

\section{Physical observables and the $\ell\to\infty$ zero curvature limit }

The principle underlying our construction of deS scattering amplitudes is the requirement that they reproduce
standard Minkowski amplitudes  in the limit of zero curvature, that is $\ell\to\infty$. This requirement will be completely sufficient to resolve several well-known ambiguities. This limit will be reached, however, not by ``flattening'' the geometry but by considering  ``high-energy'' scattering processes  probing geometry at  distances much shorter than $\ell$. The states of free particles will be classified according to the  representation theory  of de Sitter $SO(1,d)$ symmetry group, with the generators becoming quantum operators associated to physical observables. Single particle states will be assigned to unitary irreducible representations (UIRs).  $SO(1,d)$ has a lower rank than Poincar\'e algebra in the same number of dimensions, therefore the first question to address is how de Sitter observables match the familiar Minkowski observables, for instance the momentum vector. This is described by the so-called In\"on\"u-Wigner contraction \cite{ew,Enayati:2022hed}. We choose to explain it in the simplest case of $SO(1,2)$ because in higher dimensions, the contraction works in the same way as in $d=2$.

The case of deS$_2$ is relatively simple because UIRs of $SO(1,2)$ are easy to classify \cite{Sun:2021thf}. The group has rank 1 and there is only one Casimir operator
\be L^2=\frac{\hbar^2}{2}\sum_{AB}L_{AB}L^{BA}\equiv \hbar^2\Delta(1-\Delta)\label{els}\ee
 that can be constructed from the generators.\footnote{In the case of deS$_4$ universe, $SO(1,4)$ has rank 2 and there is one additional, quartic  Casimir operator, therefore the classification is more involved
\cite{Enayati:2022hed}. }
We define\footnote{In higher dimensions, one usually defines $P_\mu=\ell^{-1}L_{d\mu}$ in the natural units $\hbar=c=1$.}
\be P_0=\hbar c\ell^{-1}L_{01}\ ,\quad P_1=\hbar\ell^{-1}L_{12}\ ,\quad B=\hbar L_{20}\ .\ee
Then the algebra (\ref{alg}) reads
\be [P_0,P_1]=-i\hbar c\ell^{-2}B\ ,\quad [B,P_0]=i\hbar c P_1,\quad [B,P_1]=i\hbar c^{-1}P_0\ .\ee
Note that the factors of $\hbar$ and $c$ have been introduced in such a way that $P_0$ and $P_1$ have the dimensions of energy and momentum, respectively. As usual, the classical limit corresponds to $\hbar=0$.
In the limit of $\ell\to\infty$,   deS$_2$ algebra contracts to two-dimensional Poincar\'e algebra with the boost operator identified as $B$ and the energy-momentum vector $(P_0,P_1)$.
Since
\be P_0^2-c^2P_1^2+
\ell^{-2}c^2B^2= \ell^{-2}c^2L^2\equiv m^2c^4, \label{ons} \ee
and the term involving the $B$ operator  drops out from the l.h.s.\ when $\ell\to\infty,$  we also obtain the on-shell condition for a free particle with mass $m$.

It is interesting to look at Eq.(\ref{ons}) from the perspective of the physical universe. The accelerating expansion can be attributed to the cosmological constant $\Lambda\approx 10^{-52}$m$^{-2}$ \cite{Peebles:2002gy}. Using this value as a reference point for the curvature, the corresponding mass scale  $\hbar c\sqrt{\Lambda}\approx \hbar c/\ell\sim 10^{-33}$eV. The only UIRs of de Sitter group that allow arbitrarily large values of the quadratic Casimir operator $L^2$ are the principal series representations with $\Delta=\frac{d-1}{2}+i\mu$ and their higher spin extensions. To match the physical mass scales of order electronvolts, the parameter $\mu$ must be very large, of order $10^{33}$.

As mentioned before, free single particle states belong to UIRs of de Sitter group. On the other hand, observers live on timelike geodesics. What do they measure? In $d=2$ the answer is simple because the states belonging to UIRs with arbitrary $L^2$ (\ref{els}) can be expressed in the ``momentum'' basis of the eigenstates of  $L_{12}=\ell\hbar^{-1}P_1$, with integer eigenvalues $n$: $|n;\Delta\rangle$. How do these states evolve in time? 

The Hamiltonian operator should reflect the energy of a particle and depends on the reference frame. Cacciatori et al. \cite{Cacciatori:2007in} argued that, up to a constant factor, $\sum_{AB}K_{AB}^{(1)}K^{(2)AB}$ (with $K_{AB}$  given in Eq.(\ref{kab})) is the energy of a particle propagating on geodesic number 2  as measured in the comoving reference frame on geodesic number 1.  To that end, they choose observer 1 ``at rest at the origin $X=(0,1,0.\dots)$'' of the aforementioned vertical geodesic line and choose the vectors $v_2$ and $w_2$ describing a particle (with mass $m$) on a geodesic trajectory passing the origin but propagating with arbitrary velocity. They showed that in conformal, planar as well as in  static coordinates \cite{Spradlin:2001pw},  $mc^2\sum_{AB}K_{AB}^{(1)}K^{(2)AB}$ yields a conserved quantity associated to the time-translational invariance of the action of a particle propagating in the background of de Sitter metrics. They also showed that in the flat limit of $\ell\to\infty$, this quantity coincides with the relativistic energy of the particle.

To develop quantum field theory, the classical observer, living on a geodesic world-line specified by the vectors
 $v$ and $v$, starts from the energy proposed by Cacciatori et al.\ \cite{Cacciatori:2007in} and follows the established lore of quantum theory by replacing the Noether charge $K^{(2)AB}$ of particle's geodesics by the corresponding quantum operator $L^{AB}$. In this observer's frame, the Hamiltonian operator of a free particle is given by
\be H_0=2\sum_{AB}v_{A}w_{B}L^{AB}\  ,\ee
where $v$ and $w$ are the vectors defining observer's geodesic. For the vertical observer,
\be v=\frac{1}{2}(1,1,0)\ ,\quad w= \frac{1}{2}(1,-1,0)\ ,\ee
therefore the free Hamiltonian, \be H_0=L_{01}\ ,\label{hfr}\ee
is identified with the boost operator.
 Indeed, $L_{01}=\ell(\hbar c)^{-1}P_0$ generates boosts in the direction of $X_1$, hence it generates proper time translations without affecting the directions perpendicular to the geodesic plane. In higher number of dimensions, the same Hamiltonian governs time evolution on vertical geodesics but there are more quantum numbers characterizing UIRs \cite{Enayati:2022hed,Penedones:2023uqc} and there is a larger set of commuting operators.

\section{Quantization of scalar fields in deS$_2$}
In this section, we proceed with the canonical quantization of a free scalar field. Here again, we limit our discussion to deS$_2$ because it is technically simple, but it contains all  ingredients necessary for  higher dimensions. From now on, we set  $\hbar=c=\ell=1$.

{}{For our purposes, it is most convenient to use the global conformal coordinates $(t,\varphi)$ \cite{Sun:2021thf}, which are related to coordinates of the embedding space in the following way:
\be X^0=\tan t\ ,\quad X^1=\frac{\sin\varphi}{\cos t}\ ,\quad
X^2=-\frac{\cos\varphi}{\cos t}\  ,\label{coord}\ee\be -\frac{\pi}{2}\leq t\leq \frac{\pi}{2}\ ,\qquad 0\leq\varphi < 2\pi\ .\ee
The deS$_2$ metric is given by
\be ds^2=\frac{-dt^2+d\varphi^2}{\cos^2t}\ .\ee
\vspace*{-1cm}
\begin{figure}\centering\vspace*{-2cm}\includegraphics[scale=0.25,page=1]{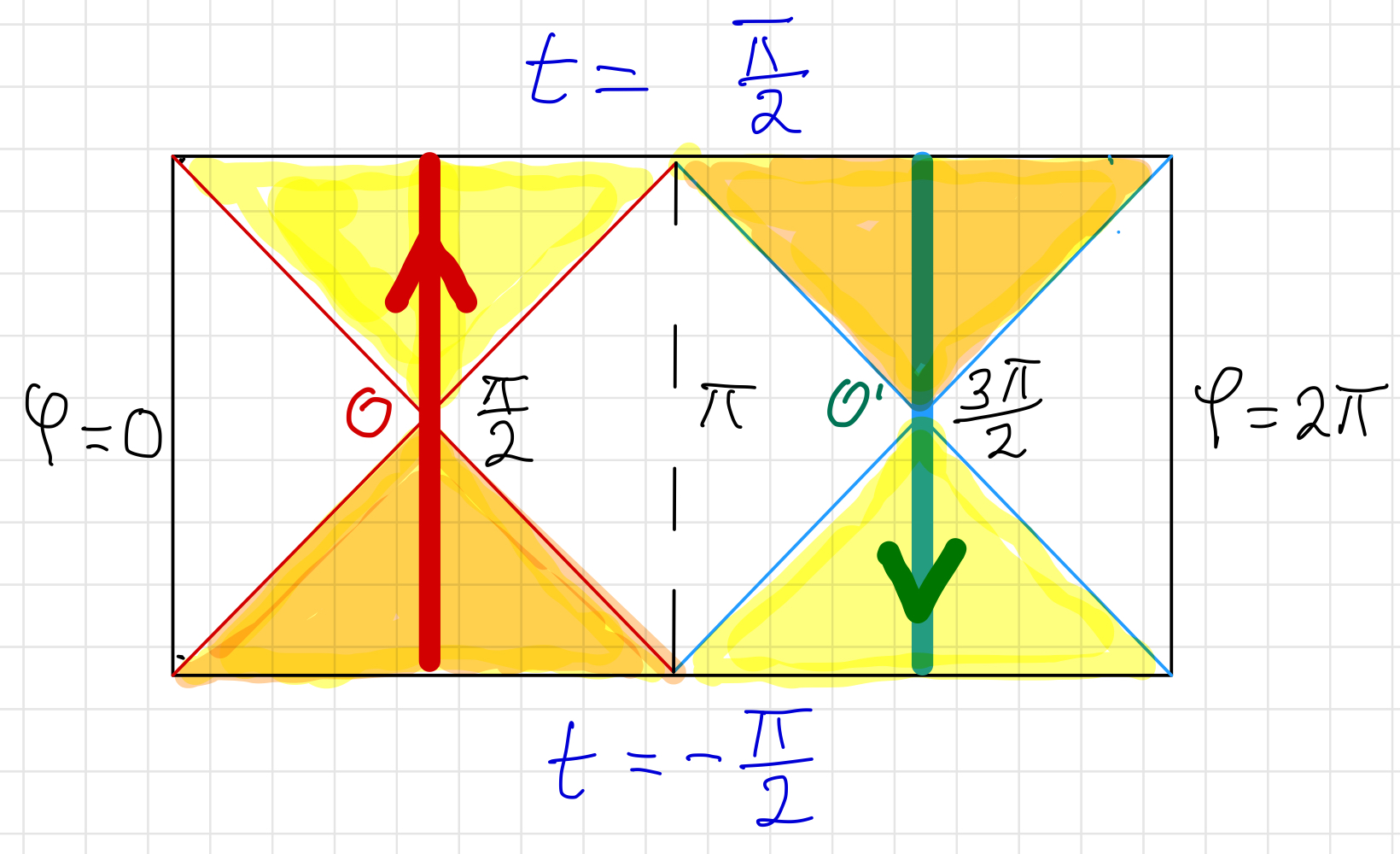}\caption{Penrose diagram for deS$_2$.}\end{figure}
\vskip 1 cm\noindent
The Penrose diagram is shown in Fig.\ 3. The world line of the observer $\cal O$, at $\varphi=\pi/2$,  is marked in red, together with his/her/their past and future light cones originating at $t=\tau=0$. The past and future horizons enclose $\varphi\in (0,\pi)$. We also marked in green the antipodal mirror geodesic at $\varphi=3\pi/2$, which is cut by the same defining plane on the other side of deS$_2$. The antipodal observer ${\cal O}'$ is a mirror image of $\cal O$ under $X_A\to -X_A$ and his/her/their  horizons enclose $\varphi\in (\pi,2\pi)$.
Note that along these
geodesics
\be X_0(\tau)=\tan t=\pm\sinh\tau\label{xoo}\ee
for red and green, respectively.

In global cordinates (\ref{coord}), deS$_2$ symmetry is generated by the Killing vectors
\begin{align}
&\xi_1=\cos t\sin\varphi\,\partial_t+\sin t\cos\varphi\,\partial_{\varphi}= i{L}_{01}\  ,\label{kil1}\\[1mm]
&\xi_2=-\cos t\cos\varphi\,\partial_t+\sin t\sin\varphi\,\partial_{\varphi}=i{ L}_{02}\label{kil2}\\[1mm]
&\xi_3=-\partial_{\varphi}=i{ L}_{12}\ ,\label{kil3}
\end{align}
Note that
\be L^2=\nabla^2\ ,\ee
where $\nabla^2$ is the two-dimensional Laplace operator.
Along the vertical geodesics, $\xi_1$ of Eq.(\ref{kil1}) translates in proper time:
\be \xi_1\big(\varphi=\frac{\pi}{2}\big)=\cos t\,\partial_t=\frac{d}{d\tau}\ ,\quad  \xi_1\big(\varphi=\frac{3\pi}{2}\big)=-\cos t\,\partial_t=\frac{d}{d\tau}\ ,\label{ki}\ee
where at $\varphi=\frac{\pi}{2}$, $\tau$ refers to the proper time on the red geodesics, while  at $\varphi=\frac{3\pi}{2}$ it refers to the proper time on the green one, which runs in the opposite direction, see Eq.(\ref{xoo}),

The Lagrangian density of the free real scalar field $\phi(t,\varphi)$ is
\be{\cal L}=\frac{\sqrt{-g}}{2}(-g^{\mu\nu}\partial_\mu\phi\partial_\mu\phi-m^2\phi^2)=\frac{1}{2}\Big[(\partial_t\phi)^2-
(\partial_\varphi\phi)^2-\frac{m^2}{\cos^2t}\phi^2\Big]\  .
\ee
The corresponding Klein-Gordon equation reads
\be (\nabla^2-m^2)\phi=(L^2-m^2)\phi=0\  .\label{mot}\ee
{}For the reasons explained before, we are interested in the principal series representations,\footnote{In $d=2$, the dimension of principal series $\Delta=\frac{1}{2}+i\mu$ and $m^2=\frac{1}{4}+\mu^2$. In Section 6, we will make a precise connection between one-particle states and principal series representations.} therefore we assume \be m^2\ge\frac{1}{4}\ .\ee
After we expand the field in Fourier modes labeled by integer $n$,
 \be \phi(t,\varphi)=\sum_{n=-\infty}^{n=\infty}\phi_n(t)e^{in\varphi}\  ,\label{wavel}\ee
we obtain
\be\frac{d^2\phi_n}{dt^2} +n^2 \phi_n + \frac{m^2}{\cos^2t} \phi_n=0\ .\label{kg3}\ee
The solutions are linear combination of two basis functions that can be expressed in terms of the Ferrers' (associated Legendre) functions,
\be
\phi_{1n}(t)= \sqrt{\cos t} \, P_{n-\frac{1}{2}}^{\,i\mu}(\sin t) \, ,\qquad
\phi_{2n}(t) =  \sqrt{\cos t} \, Q_{n-\frac{1}{2}}^{\,i\mu}(\sin t) \, .
\label{kgsol}\ ,\ee
with \be\mu=\pm\sqrt{m^2-\frac{1}{4}}\ . \ee
Both functions are symmetric under $n\to -n$, but the replacement $\mu\to-\mu$ yields different solutions.

To quantize this theory, we follow the same steps as in  Minkowski spacetime. We need to identify the ``positive energy/frequency'' modes which behave as standard plane waves in the limit of $n\to\infty$ {\it i.e}.\ in the limit of short wavelengths (see Eq.(\ref{wavel})), in which  they probe de Sitter spacetime at short distances.
 To that end, we examine the asymptotic expansions
\ba
\phi_{1n}(t)\sim \sqrt{\frac{2}{\pi}} \left(| n|-\frac{1}{2}\right)^{i\mu-\frac{1}{2}}  \cos\big[ |n|(t-\frac{\pi}{2})+\frac{\pi}{2} (\frac{1}{2}-i\mu)\big]\ , \\
\phi_{2n}(t)\sim \sqrt{\frac{\pi}{2}} \left(| n|-\frac{1}{2}\right)^{i\mu-\frac{1}{2}}  \sin\big[ |n|(t-\frac{\pi}{2})+\frac{\pi}{2} (\frac{1}{2}-i\mu)\big]\ ,
\ea
with the remainders  suppressed by factors of order $|n|^{-1}$. The above expansions are valid in the whole range of $t$, that it throughout entire observers' history, except near the end points $t=\pm\frac{\pi}{2}$ {\it i.e}.\ $\tau\to\pm\infty$. We can construct the following combinations:
\begin{align}
\phi_{n}(t)_\pm &=\sqrt{\cos t}\, P^{\mp i\mu}_{n-\frac{1}{2}}(\sin t) \mp \frac{2i}{\pi} \sqrt{\cos t}\, Q^{\mp i\mu}_{n-\frac{1}{2}}(\sin t)\label{pw}\\[1mm] &~~~~~\sim\, e^{\mp i |n|t} \,\sqrt{\frac{2}{\pi}} e^{
\mu\pi\over 2}\left(| n|-\frac{1}{2}\right)^{\mp i\mu-\frac{1}{2}}  e^{\pm i\frac{\pi}{2} (|n|-\frac{1}{2})}\ ,\nonumber
 \end{align}
which behave as  positive and negative frequency waves, respectively, in global time coordinate $t$. de Sitter symmetry transformations of the geodesic observers correspond to (proper) Lorentz transformations in the embedding space, therefore they do not reverse the course of their proper time and the signs of  frequencies remains the same in all transformed reference frames. There is no mixing between positive and negative frequency modes induced by proper (continuous) symmetry transformations.

By using the properties of associated Legendre functions, one can show that $\phi_n=\phi_{-n}$ and that under complex conjugation, $\phi_{ n}(t)_+^*=\phi_{ n}(t)_-$. The observers located near the hyperboloid throat at $X_0=0$, where $t\approx\pm\tau$ on vertical geodesics, detect them as positive and negative frequency oscillations in their own proper time.  For ${\cal O}$ and ${\cal O}'$,  $\phi_{ n}(t)_+$ are propagating forward and backward in proper time, respectively, while  $\phi_{ n}(t)_-$  are propagating  in the opposite proper time directions.

The next step is to identify orthonormal  frequency modes, normalized  with respect to the Klein-Gordon scalar product.  In global coordinates, it reads
\be
(\phi,\phi') = i\int_0^{2\pi} d\varphi (\phi^*\partial_t \phi' -\partial_t\phi^* \phi')\  ,\label{spr}
\ee
integrated along a circle of constant $t$.
The normalized solutions have the form
\be\phi_{ n}(t,\varphi)_{\pm}=c(n,\mu)_\pm\, \phi_{n}(t)_\pm e^{\pm i n\varphi}.\label{mno} \ee
with $\phi_{n}(t)_\pm$ written in Eq.(\ref{pw})
and the normalization constants $c(n,\mu)_\pm$ determined by requiring
\be (\phi_{ m+},\phi_{ n+})=(\phi_{ m-},\phi_{ n-})=\delta_{mn}\ ,\quad (\phi_{ m+},\phi_{ n-})=0\ ,
\ee
The most convenient choice of time slice for computing the scalar products (\ref{spr}) is $t=0$ because the values of associated Legendre functions and their derivatives at this point are given by relatively simple expressions involving Euler's Gamma functions. In this way, we obtain
\be c(n,\mu)_+=c(n,\mu)_-=\frac{e^{-{\mu\pi\over 2}}}{2\sqrt 2}\ .\label{cnor}\ee
By using the properties of associated Legendre functions, the properly normalized modes (\ref{mno}) can be also expressed in terms of the conical (Mehler) functions:
\begin{align}
\phi_{ n}(t,\varphi)_{+} & =\frac{\Gamma(n+\frac{1}{2}-i\mu)}{2\sqrt{\pi}} P^{-n}_{-\frac{1}{2}-i\mu}( i \tan t)  \, e^{ i n\varphi}\ ,\nonumber\\
\phi_{ n}(t,\varphi)_{-} & =\frac{\Gamma(n+\frac{1}{2}+i\mu)}{2\sqrt{\pi}} P^{-n}_{-\frac{1}{2}-i\mu}(-i \tan t)  \, e^{- i n\varphi}\ .\label{modes}
\end{align}
Here again,  $\phi_{ n}(t,\varphi)_+^*=\phi_{ n}(t,\varphi)_-$. Both positive and negative frequency modes are symmetric under $n\to -n$, what can be shown by using the properties of Legendre functions. Furthermore, since conical functions are symmetric under $\mu\to -\mu$, the functions $\phi_{ n}(t,\varphi)_{\pm}$ aquire phase factors upon such transformations. Note that the argument of the wavefunctions (\ref{modes}), $i\tan t= iX_0$, is the Wick-rotated time coordinate of the embedding Minkowski spacetime. Actually, these de Sitter harmonics can be obtained from Euclidean spherical harmonics by a Wick rotation  \cite{Mottola:1984ar,Allen:1985ux}; the angular momentum is identified as $-\Delta$.

We conclude that the scalar field can be expanded as
\be \phi(t,\varphi)=\sum_{n=-\infty}^{n=\infty}\Big(a_n\phi_{ n}(t,\varphi)_{+}+a^\dagger_n\phi_{ n}(t,\varphi)_{-}\Big)\ .
\label{fiel}\ee
In quantum theory, the creation and annihilations operators satisfy the commutation relations
\be[a_m,a_n^\dagger]=\delta_{mn}\, ,\qquad [a_m,a_n]=[a_m^\dagger,a_n^\dagger]=0\ .\label{crel}\ee
The vacuum state is annihated by all annihilation operators:
\be a_n|0\rangle=0\ .\ee
This vacuum is {\it unique}, de Sitter invariant and common to all observers. Since it is related to Euclidean spherical harmonics, it is usually  called  Euclidean, a.k.a.\ Bunch-Davies vacuum \cite{Mottola:1984ar,Allen:1985ux}. From our perspective it appears as the only vacuum 
associated to the  unique linear combinations of Klein-Gordon solutions (\ref{modes})
with the correct asymptotic behavior (\ref{pw}) at $n\to\infty$. One-particle states are obtained by acting on this vacuum with the  creation operators:
\be |n,\mu\rangle =a_n^\dagger|0\rangle .\label{onep}\ee
These states are orthonormal:
\be \langle m,\mu |n,\mu\rangle=\delta_{mn}\ ,\ee
which follows from the commutation relations (\ref{crel}).
 In the Section 6, we will show that they belong to the principal series representations of de Sitter group.

\section{Spectral decomposition of wavefunctions}
In this section, we discuss two-dimensional wavefunctions written in Eqs.(\ref{modes}).  We want to decompose them into Fourier modes with the frequencies conjugate to the variable
\be T \equiv \tan t= X_0\ ,\label{cyl}\ee
that is the time coordinate in embedding spacetime. We will show that in the high frequency/short wavelength limit, these frequencies determine particle energies in accordance with the usual on mass shell constraints, {\it i.e}. relativistic dispersion relations. The wavefunctions are labelled by $\mu$, which is related to the mass ($m^2=\frac{1}{4}+\mu^2$), and integer momentum $n$. As explained in Section 3, we are interested in the limit of very large $\mu$ and/or $n$ characterizing observable particles. The asymptotic behavior of conical functions for large values of these parameters is known, although the formulas describing the simultaneous limit of $\mu\to\infty, ~n\to \infty$ with  $n/\mu$ fixed are rather complicated; in this case, we will rely on numerical estimates.
\subsection{Nonrelativistic limit $n\ll \mu\to \infty$}
In the $n\ll \mu\to \infty$ limit, the conical functions can be approximated near $T=0$  by
\be P^n_{-\frac{1}{2}-i\mu}(iT)\approx P^0_{-\frac{1}{2}-i\mu}(iT)\approx\frac{\sqrt{\pi}}{\Gamma\left( \frac{3}{4} +\frac{i\mu}{2}\right) \Gamma\left( \frac{3}{4} -\frac{i\mu}{2}\right)}\,e^{-i|\mu| T}\quad (n\ll |\mu|\to \infty).
\ee
{}For large $\mu$, the Gamma functions can be approximated by using  Stirling's formula for complex numbers.
The corresponding wavefunctions (\ref{modes}) read
\be \phi_n(T,\varphi)_+ \approx \frac{e^{-i|\mu| T}e^{in \varphi}}{\sqrt{2|\mu|}\sqrt{2\pi}}\
\qquad (T\approx 0,\,\makebox{as}~|\mu|\to\infty)\ ,\label{nonr}\ee
modulo constant phase factors. This means that Fourier transforms should peak at $\omega=|\mu|$, while the fluctuations outside the neighborhood of $T=0$ should be visible at lower frequencies.

We will decompose conical functions into Fourier modes by using the following integral representation:
\be P^0_{-\frac{1}{2}-i\mu}(iT)=\sqrt{2\over \pi} \,\frac{1}{\Gamma\left( \frac{1}{2} +i\mu\right) \Gamma\left( \frac{1}{2}-i\mu\right)}\int_0^\infty t^{-\frac{1}{2}}K_{i|\mu|}(t)e^{-itT}dt\ ,
\ee
where $K_{i\mu}$ is the modified Bessel function of the second kind. Then the Fourier transform
 \be\frac{1}{2\pi}\int_{-\infty}^{\infty} P^0_{-\frac{1}{2}-i\mu}(iT)\, e^{i\omega T}dT
=\frac{2}{\sqrt{\pi}\Gamma\left( \frac{1}{2} +i\mu\right) \Gamma\left( \frac{1}{2}-i\mu\right)}\frac{K_{i|\mu|}(\omega)}{\sqrt{2\omega}}\ .
\ee
In this way, we obtain the spectral decomposition:
\be \widehat{\!\phi}_n(\omega,\varphi)_+\equiv\frac{1}{2\pi}\int_{-\infty}^{\infty}\phi_n(T,\varphi)\, e^{i\omega T}dT=\rho_\mu(\omega)e^{in\varphi},\label{ftr}\ee
with the spectral density
\be \rho_\mu(\omega)=\frac{1}{\sqrt{2\omega}\sqrt{2\pi}}\,e^{|\mu|\pi\over 2}\pi^{-1}K_{i|\mu|}(\omega)\qquad (\makebox{as}~|\mu|\to\infty)\ ,
\ee
modulo a constant phase factor. Anticipating a peak at $\omega=|\mu|$, we write the argument of the Bessel function as
\be K_{i|\mu|}(\omega)=K_{i|\mu|}(|\mu|z)\ , \qquad z\equiv\frac{\omega}{|\mu|}\ ,\ee
In the large $|\mu|$ limit, we can use the asymptotic expansion \cite{olv,bal}:
\be K_{i|\mu|}(|\mu|z)\approx\pi\, e^{-|\mu|\pi\over 2}{\rm R}_\mu(z)\ ,\ee
with
\be {\rm R}_\mu(z)=|\mu|^{-\frac{1}{3}}\bigg( {4\xi\over 1-z^2}\bigg)^{1\over 4}\makebox{Ai}(-|\mu|^{2\over 3}\xi)\ ,\label{rdef}\ee
where Ai is the Airy function and the variable $\xi$ is defined through
\be\frac{2}{3}\xi^{3\over 2}(z)=\ln\bigg({1+\sqrt{1-z^2}\over z}\bigg)-\sqrt{1-z^2}\ .\label{xid}\ee
The above formula is valid for $z\le 1$. {}For $z\ge 1$ a similar ``Airy expansion'' can be found in Refs. \cite{olv,bal}, hence ${\rm R}_\mu(z)$ can be extended to all $z>0$.
Note that the spectral density
\be \rho_\mu(\omega)=\frac{{\rm R}_\mu(z)}{\sqrt{2\omega}\sqrt{2\pi}}\ ,~~~\qquad (\makebox{as}~|\mu|\to\infty)\ .
\ee
\begin{figure}\centering\includegraphics[scale=0.45,page=1]{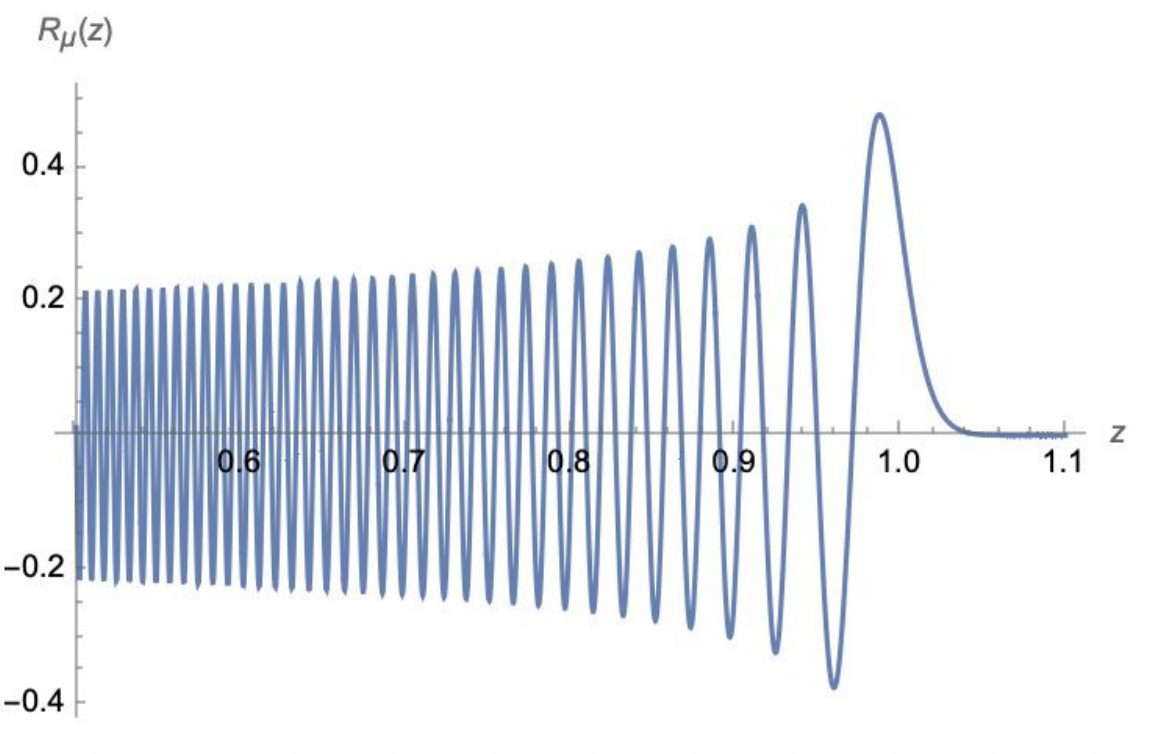}\vskip 0 cm\caption{${\rm R}_\mu(z)$ for $\mu=500$.}\vskip -0 cm\end{figure}

As an example, we plot ${ \rm R}_{\mu}(z)$ in Fig.4, for $\mu=500$. {}For $z<1~(\omega<|\mu|)$, it is an oscillating function with increasing amplitude. It does indeed peak at $z\approx 1~ (\omega\approx |\mu|) $, which is expected to be the highest frequency in the spectrum, and then falls off rapidly. This peak becomes even more dramatic if one takes the formal limit of $\mu\to\infty$, by using
\be\lim_{\epsilon\to 0}\frac{1}{\epsilon}
\makebox{Ai}\Big({ x\over\epsilon}\Big)=\delta(x)\ .\label{ail}\ee
The function $\xi(z)$, defined in \ Eq.(\ref{xid}), has a single zero at $z=1$ with $\xi'(1)=\sqrt[3]{2}$.
Then it follows from Eqs. (\ref{rdef}) and (\ref{ail})  that
\be \lim_{\mu\to\infty}{\rm R}_\mu(z)=\frac{1}{|\mu|}\delta (1-z)=\delta(\omega-|\mu|)\  ,\ee
so that
\be\rho_\mu(\omega)~\to~\frac{\delta(\omega-|\mu|)}{\sqrt{2\omega}\sqrt{2\pi}}\ .\ee
After taking the inverse Fourier transform of Eq.(\ref{ftr}), we recover the nonrelativistic wavefunction of Eq.(\ref{nonr}).

We conclude that in de Sitter space, nonrelativistic wavefunctions  contain a wide range of frequencies when particles have the mass parameter $\mu$ comparable to the curvature/cosmo\-logical scale. Due to the rapid oscillations of spectral density, it appears that the frequencies are quantized.  The spectrum narrows for heavier particles and peaks at the frequencies corresponding to the physical mass in flat spacetime. 
\subsection{Ultrarelativistic limit  $\mu\ll n\to \infty$}
We already know from section 4, {\it c.f}.\ Eqs.(\ref{pw}) and (\ref{cnor}), that in this limit,
\be \phi_n(t,\varphi)_+ \approx \frac{e^{-i|n| t}e^{in \varphi}}{\sqrt{2|n|}\sqrt{2\pi}}  =\Big(\frac{1-iT}{1+iT}\Big)^{\frac{|n|}{2}}\!
\frac{e^{in \varphi}}{\sqrt{2|n|}\sqrt{2\pi}} \
\qquad (T\approx 0,\,\makebox{as}~|n|\to\infty)\ ,\label{ulnr}\ee
modulo constant phase factors.
In the ultrarelativistic limit, we write its Fourier transform as
\be \widehat{\!\phi}_n(\omega,\varphi)_+\equiv\frac{1}{2\pi}\int_{-\infty}^{\infty}\phi_n(T,\varphi)\, e^{i\omega T}dT=\rho_n(\omega)e^{in\varphi},\label{ftr2}\ee
After performing the Fourier integral, we obtain
\be \rho_n(\omega)=\frac{2}{\sqrt{2|n|}\sqrt{2\pi}}e^{-\omega} L^1_{|n|/2-1}(2\omega)\ ,\ee
modulo a constant phase factor. Here, $L^1_{|n|/2-1}$ is the Laguerre function. In order to zoom on the region $\omega/|n|\equiv z\approx 1$, we can use the asymptotic expansion of  Laguerre function at $\omega\to\infty$.  As a result, we obtain
\be \rho_n(\omega)=\frac{{\rm R}_n(z)}{\sqrt{2|n|}\sqrt{2\pi}}\ ,~~~\qquad (\makebox{as}~|n|\to\infty)\ ,
\ee
where ${\rm R}_n(z)$ can be expressed in terms of the Airy function, in a similar way as ${\rm R}_\mu(z)$ in Eq.(\ref{rdef}). As an example, we plot ${\rm R}_n(z)$ in Fig.5, for $n=500$. We see that the frequency pattern is very similar to the nonrelativistic case. Furthermore, we can show that, as expected,
\be \lim_{n\to\infty}{\rm R}_n(z)=\delta(\omega-|n|)\  ,\ee
therefore
\be\rho_n(\omega)~\to~\frac{\delta(\omega-|n|)}{\sqrt{2|n|}\sqrt{2\pi}}\ .\ee
\begin{figure}\centering\includegraphics[scale=0.45,page=1]{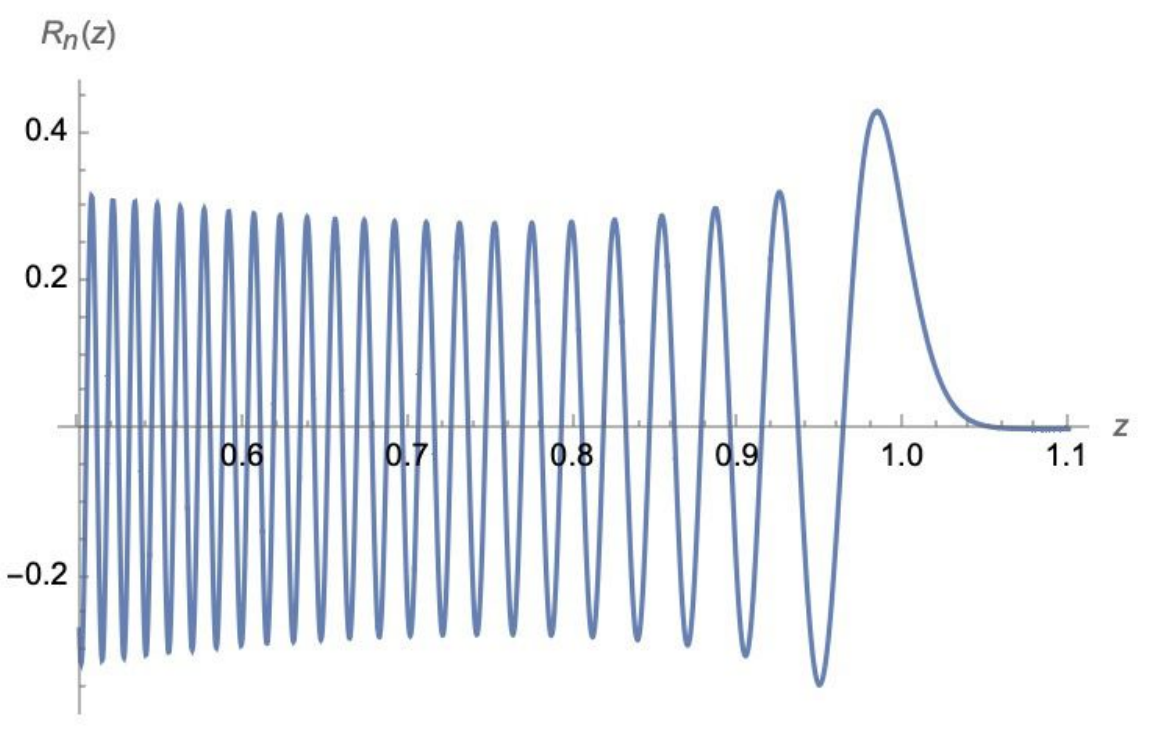}\vskip 0 cm\caption{${\rm R}_n(z)$ for $n=500$.}\vskip -0 cm\end{figure}

A word of caution is here in order. The computations of spectral densities were based on the asymptotic expansions of wavefunctions which are valid everywhere except at very large values of $T=X_0$. They fail in the asymptotic past and future, {\em i.e}.\ near $t=\pm\frac{\pi}{2}$. For that reason, the corresponding spectral densities cannot be trusted at very low frequencies, near $z= 0$. Actually, one expects more suppression at low frequencies than shown in Figs.\ 4 and 5. In the ultrarelativistic case $n\to\infty$, however, there exist so-called uniform expansions \cite{Jones} which are valid in the lower $(t<0)$ and upper $(t>0)$ half of the hyperboloid, separately. For $0<t<\pi/2$, we define $t_P=t-\frac{\pi}{2}$. Then \cite{Jones}:
\be\phi_n(t,\varphi)_+\approx e^{-\frac{\mu\pi}{2}}\sqrt{-t_P} \,H_{i\mu}^{(1)}(-|n|t_P)\,e^{in\varphi},\label{hank}\ee
where $H_{i\mu}^{(1)}$ is the Hankel function of the first kind. This asymptotic formula holds for all $X_0\geq 0$, in particular on the ``Poincar\'e patch'' of  $X^0> -X^1$. 

\subsection{Special relativistic limit  $\mu\to\infty,~ n\to \infty$ with fixed $n/\mu$}
In this limit, the asymptotic expansions of conical functions are known \cite{dunst}, although they are given by rather complicated expressions. From Eq.(\ref{kg3}), we know that near $t=0$,  the positive frequency solution has $\omega=\sqrt{n^2+m^2}$. For not too large values of the parameters $n$ amd $\mu$, the frequency pattern can be studied numerically. Spectral densities do indeed peak at $\omega=\sqrt{n^2+m^2}$. The peaks become more narrow for larger parameters. All three limits  of the ``positive frequency'' wavefunctions share the same features:
\begin{itemize}
\item The spectral density $\rho(\omega)$ is real
\item $\rho(\omega)=0$ for $\omega< 0$
\item $\rho(\omega)$  oscillates rapidly in the region between $\omega=0$ and $\omega=\sqrt{n^2+m^2}$. It peaks near $\omega=\sqrt{n^2+m^2}$, which in the flat limit corresponds to the energy $E=\omega$ and momentum $P=n$ related by the on-shell condition $E^2-P^2=m^2$. In the limit  of $\mu\to\infty,~ n\to \infty$,
\be\rho(\omega)~\to~\frac{\delta(\omega-E)}{\sqrt{2\omega}\sqrt{2\pi}}\ .\ee
with $E=\sqrt{n^2+m^2}$.
\item $\rho(\omega)$ is exponentially suppressed for $\omega>E$.
\end{itemize}

In comparison with the flat case, de Sitter wavefunctions describe  wavepackets with the frequency spectrum determined by geometry.
The frequency spread shrinks as  the masses and/or momenta become larger than the curvature scale. In this limit, the frequencies and wavelengths are related by the usual relativistic dispersion relations.

\section{The Hamiltonian and symmetry operators on Hilbert space}
In scalar field theory with canonical kinetic energy terms and arbitrary potential
\be V(\phi)=\frac{1}{2}m^2\phi^2+V_I(\phi)\ ,\ee the energy-momentum tensor is given by
\be T_{\mu\nu}={\partial\phi\over\partial x^\mu}{\partial\phi\over\partial x^\nu}-\frac{1}{2}g_{\mu\nu}
{\partial\phi\over\partial x^\alpha}g^{\alpha\beta}{\partial\phi\over\partial x^\beta}-V(\phi)g_{\mu\nu}\ . \ee
{}For each symmetry generated by a Killing vector $\xi^\mu$, there exists a corresponding Noether current
\be J_{\mu}=T_{\mu\nu}\,\xi^\nu\  .\label{tkil}\ee
On a surface of constant $t$, the conserved charge density is given by
\be J_{0}=T_{0\nu}\,\xi^\nu={\partial\phi\over\partial t}{\partial\phi\over\partial x^\nu}\xi^{\nu}
-\frac{1}{2}{\partial\phi\over\partial x^\alpha}g^{\alpha\beta}{\partial\phi\over\partial x^\beta}\xi_0
-V(\phi)\xi_0\ .\label{dens}
\ee

In deS$_2$, there are three Killing vectors, (\ref{kil1})-(\ref{kil3}). In free scalar theory, the corresponding charge operators, with the densities (\ref{dens}) integrated on $t=0$ circular slices, are given by
\begin{align} \label{q1}Q_1& =\frac{1}{2}\int_0^{2\pi}d\varphi\,\sin\varphi\, :\Big[\Big( \frac{\partial \phi}{\partial t}\Big)^2
+\Big( \frac{\partial \phi}{\partial\varphi}\Big)^2+m^2\phi^2\Big]:\ ,\\
Q_2&=-\frac{1}{2}\int_0^{2\pi}d\varphi\,\cos\varphi\,:\Big[\Big( \frac{\partial \phi}{\partial t}\Big)^2
+\Big( \frac{\partial \phi}{\partial\varphi}\Big)^2+m^2\phi^2\Big]:\ ,\\[1mm]
Q_3&=-\int_0^{2\pi}d\varphi\,:\frac{\partial \phi}{\partial t}\frac{\partial \phi}{\partial\varphi}:\ ,
\end{align}
where  we use normal ordering and  $\phi\equiv \phi(0,\varphi)$. The values of  fields and their time derivatives at $t=0$ can be obtained from Eqs.\
(\ref{modes}) and (\ref{fiel})
by using well-known ``special values'' of associated Legendre functions:
\begin{align}
&P^{-n}_{-\frac{1}{2}-i\mu}(i\tan t)\big|_{t=0} = \frac{2^{-n} \sqrt{\pi}}{\Gamma\left( \frac{3}{4} +\frac{n-i\mu}{2}\right) \Gamma\left( \frac{3}{4} +\frac{n+i\mu}{2}\right)} \, ,\\
&\frac{d P^{-n}_{-\frac{1}{2}-i\mu}(i\tan t)}{dt}\Big|_{t=0} =-i \frac{2^{-n+1}\sqrt{\pi}}{\Gamma\left( \frac{1}{4} +\frac{n-i\mu}{2}\right) \Gamma\left( \frac{1}{4} +\frac{n+i\mu}{2}\right)} \, .
\end{align}
After tedious computations, often using
\be\textstyle
\Big|\Gamma\left(\frac{3}{4}+\frac{n-i\mu}{2}\right)\Gamma\left( \frac{1}{4}+\frac{n-i\mu}{2}\right)\!\Big|^2=2^{1-2n}\pi\, \Big|\Gamma\left( \frac{1}{2} +n-i\mu\right)\! \Big|^2\label{eq:4Gammaproduct}\ ,
\ee
which follows from  $\Gamma(z)\Gamma\left(z+\frac{1}{2}\right)=2^{1-2z}\sqrt{\pi}\,\Gamma(2z)$, we obtain
\begin{align}
&Q_3 =\sum_{n=-\infty}^{n=\infty}n \,a^\dagger_n a_n\equiv L_{12}\ ,\\
&Q_+ =-Q_2+iQ_1=\sum_{n=-\infty}^{n=\infty}(n+
\Delta) \,a^\dagger_{n+1} a_n\equiv L_+\ ,\\
& Q_- =-Q_2-iQ_1=\sum_{n=-\infty}^{n=\infty}(n-
\Delta) \,a^\dagger_{n-1} a_n\equiv L_-\  ,
\end{align}
with \be \Delta=\frac{1}{2}+i\mu\ .\ee
Acting on one-particle states (\ref{onep}),
\be
L_{12}|n,\mu\rangle=n\, |n,\mu\rangle\ ,\qquad
 L_{\pm}|n,\mu\rangle=(n\pm\Delta)\, |n\pm 1,\mu\rangle\ ,\label{all}
\ee
which shows that $|n,\mu\rangle$ form a principal series representation of $SO(1,2)$ with dimension $\Delta=\frac{1}{2}+i\mu$. Unlike in Minkowski spacetime, the momenta $P_1=L_{12}=n$ are quantized because the spacelike dimension is compact.

Following the reasoning outlined in Section 3, we define the Hamiltonian operator as the charge associated to the Killing boost vector $\xi_1$ (\ref{kil1}), with the charge density written in Eq.(\ref{dens}). Indeed, in free scalar theory,  $H_0= Q_1=L_{01}$ (\ref{q1}), as anticipated in Eq.(\ref{hfr}). In the interacting theory, the density (\ref{dens}) (specified to the boost charge) yields the Hamiltonian
\be H=H_0+H_I\ ,
\label{h1}\ee
with the interaction Hamiltonian
\be H_I= \int_0^{2\pi}d\varphi\,\sin\varphi\, :H_I[\phi(0,\varphi)]:\label{hint}\ee
Note that the Hamiltonian density is positive for $\varphi $ in the interval $(0,\pi)$, {\it i.e}.\ within the horizon of  observer $\cal O$. It is negative, however, for $\varphi$ in the interval $(\pi,2\pi)$, {\it i.e}.\ within the horizon of antipodal observer ${\cal O}'$. This is expected, because the Hamiltonian generates time evolution in {\it proper\/} time which, as explained below Eq.(\ref{ki}),  runs for  ${\cal O}'$ in the direction opposite to $\cal O$. This Hamiltonian is a hermitean operator.
\section{The S-matrix}
We will construct  scattering amplitudes in  Schr\"odinger picture, by using the constant Hamiltonian written in Eqs.\ (\ref{h1}) and (\ref{hint}).
We assume that the incoming particles are described by a free state $\ket{\psi}^{in}=\ket{\psi(\tau^{\prime})}$ at proper time $\tau^{\prime}\rightarrow-\infty$. Similarly, outgoing state $\ket{\phi}^{out}$ will be described by a free state  $\ket{\phi(\tau)}$ at time $\tau\rightarrow+\infty$. In Schr\"odinger picture, the transition amplitude can be evaluated at  $\tau=0$ \cite{col}. We evolve states from $\tau$ and $\tau'$ to $\tau=0$ by using the full Hamiltonian:
\be
^{out}\braket{\chi\,}{\!\psi}^{in}=\matrixel{\chi(\tau)}{e^{-iH\tau}e^{iH\tau^\prime}}{\psi(\tau^\prime)}\ .
\ee
On the other hand, the free states $\ket{\psi(\tau^{\prime})}$ and $\ket{\chi(\tau)}$ can be expressed in terms of
free states at $\tau=0$: $\ket{\psi(\tau^{\prime})}=e^{-iH_0\tau'}\ket{\psi(0)}$,  $\ket{\chi(\tau)}=e^{-iH_0\tau}\ket{\chi(0)}$. In this way, we obtain the transition amplitude as a matrix element of an operator between free states:
\be ^{out}\braket{\chi\,}{\!\psi}^{in}
=\matrixel{\chi(0)}{e^{iH_0\tau}e^{-iH(\tau-\tau^\prime)}e^{-iH_0\tau^\prime}}{\psi(0)}=
 \matrixel{\chi(0)}{U(\tau,\tau^\prime)}{\psi(0)}\ee
 with the time evolution operator \cite{col}
 \be U(\tau,\tau^\prime)=e^{iH_0\tau}e^{-iH(\tau-\tau^\prime)}e^{-iH_0\tau}.\label{tev}\ee
The free states $\ket{\psi(0)}$ and $\ket{\chi(0)}$ belong to the representations of de Sitter group. In two dimensions, we can use the momentum basis $|n,\mu\rangle$ for single-particle states and a similar momentum basis for multi-particle states. 

In order to obtain an explicit form of the time evolution operator, we use perturbation theory, which allows expressing it as series in powers of  the interaction Hamiltonian $H_I$:
\be U(\tau\to\infty,\tau^\prime\to-\infty)=T\, \makebox{exp}\Big(-i\int_{-\infty}^{\infty}H_I(\tau)\, d\tau\Big)\ ,\label{uop}\ee
where $T$ denotes time ordering and $H_I(\tau)$ in the interaction Hamiltonian in the ``interaction'' picture:
\be H_I(\tau)=e^{iH_0\tau} H_I\,
e^{-iH_0\tau} .\ee
Eq.(\ref{uop}) is called Dyson's formula.

The operator $e^{iH_0\tau}$ boosts by rapidity $\tau$ in the $X_1$ direction of the embedding space. Acting on the scalar field in deS$_2$,
\be
e^{iH_0\tau} \phi(0,\varphi)\,
e^{-iH_0\tau} =\phi(t',\varphi')\ ,\ee
with the coordinates transformed according to
\be
\sin t'(\tau,\varphi)=\frac{\sinh\tau\sin\varphi}{\sqrt{1+\sinh^2\tau\sin^2\varphi}}\ ,\qquad \cos\varphi'(\tau,\varphi)=\frac{\cos\varphi}{\sqrt{1
+\sinh^2\tau\sin^2\varphi}}\ ,
\ee
{\it c.f}.\ Eq.(\ref{coord}).
The integral in the exponent of the time evolution operator in Eq.(\ref{uop}) can be now written as
\be\int_{-\infty}^{\infty}H_I(\tau)d\tau=\int_{-\infty}^{\infty}d\tau\int_{0}^{2\pi}d\varphi\,\sin\varphi \,H_I[\phi(t'(\tau,\varphi),\varphi'(\tau,\varphi))]\ee
At this point, it is convenient to change the integration variables from $(\tau,\varphi)$ to $(t',\varphi')$. After including the Jacobian, we obtain
\be\int_{-\infty}^{\infty}H_I(\tau)d\tau=\int_{-\infty}^{\infty}\frac{dt'}{\cos^2t'}\int_{0}^{2\pi}d\varphi' \,H_I[\phi(t',\varphi')]=\int d^2x\sqrt{-g}\,H_I[\phi(x)]\ ,
\ee
which is exactly the same volume integral as in Minkowski spacetime, but now integrated over de Sitter volume! The above result can be generalized  to $d$-dimensional deS$_d$ in a straightforward manner. We conclude that  in  deS$_d$, the  scattering amplitudes are given by the generalized Dyson's formula:
\be ^{out}\braket{\chi\,}{\!\psi}^{in}
=\matrixel{\chi(0)}{T\, \makebox{exp}\Big(-i\int d^dx
\sqrt{-g}\,H_I[\phi(x)]
\Big)}{\psi(0)}\ .\label{dys}\ee
The corresponding S-matrix is manifestly unitary. Eq.(\ref{dys}) is the main result of the present work. In the sequel \cite{tbz}, we will apply it to various examples and discuss general properties of scattering amplitudes. Here, we limit ourselves to a simple three-scalar amplitude.

\section{Example: Three-scalar amplitude in deS$_2$}
As an example of interacting theory, we consider scalar fields in deS$_2$ with the interaction potential
\be V_I(\phi)=\frac{\lambda}{3!}\phi^3\ .\label{pot3}\ee
We are interested in the amplitude  for the process $(n_1,\mu_1)\to (n_2,\mu_2)(n_3,\mu_3)$, with all particles in the principal series representation. According to Dyson's formula (\ref{dys}), at the leading order in the coupling constant $\lambda$, it is given by
\begin{align}
 ^{out}\big\langle n_2,\mu_2;n_3,\mu_3\big|n_1,\mu_1\big\rangle^{in}=\, -i\int d^2x
\sqrt{-g}\big\langle n_2,\mu_2;n_3,\mu_3\big|\!:\!V_I[\phi(x)]\!:\!\big|n_1,\mu_1\big\rangle\nonumber\\
=\, -i\frac{\lambda}{3!}\int_{-\pi/2}^{\pi/2}{dt\over \cos^2t}\int_0^{2\pi}\!d\varphi\,\big\langle n_2,\mu_2;n_3,\mu_3\big|\!:\!\phi^3(t,\varphi)\!:\!\big|n_1,\mu_1\big\rangle\ .
\end{align}
After reducing external states to the vacuum and changing the integration variables from $(t,\varphi)$ to the cylindrical (in the embedding space) coordinates $(T,\varphi)$, {\it c.f}.\ Eq.(\ref{cyl}), we obtain:\footnote{We do not want to be pedantic here, therefore we skip the $1/\sqrt{2}$ factor due to the proper normalization of two-particle states (of identical particles). Later, we do the same thing for Minkowski amplitudes.}
\begin{align}
^{out}\big\langle n_2,\mu_2;n_3,\mu_3\big|n_1,\mu_1\big\rangle^{in}= -i\lambda
\int_{-\infty}^{\infty}\!dT\int_0^{2\pi}\!d\varphi\;  \phi_{n_1}(T,\varphi)_+\phi_{n_2}(T,\varphi)_- \phi_{n_3}(T,\varphi)_-\ .
\end{align}
As expected, the amplitude is determined by the integrated overlap of the wavefunctions. It acquires a very simple form when expressed in terms of spectral densities, see Eq.(\ref{ftr}),
\begin{align} & ^{out}\big\langle n_2,\mu_2;n_3,\mu_3\big|n_1,\mu_1\big\rangle^{in}=\nonumber\\
& ~~~~~~-i\lambda(2\pi)^2\delta(n_1-n_2-n_3)\int_0^\infty\!\!  d\omega_1d\omega_2d\omega_3\, \delta(\omega_1-\omega_2-\omega_3)\,\rho_1(\omega_1)
\rho_2(\omega_2)\rho_3(\omega_3)\ ,\label{amp1}
\end{align}
where we took into account that the spectral densities of deS$_2$ wavefunctions are real.

The amplitudes (\ref{amp1}) describe the scattering processes of wave packets. Each packet has a definite momentum $n$ but the frequencies $\omega$ are distributed with the densities $\rho(\omega)$. This is different from flat spacetime, where the frequency is identified with observable energy and determined by the on-shell condition $p^2+m^2=0$, which
implies $\omega=\sqrt{n^2+m^2}$.  On general grounds, such a frequency spread follows from the cosmic uncertainty principle
$\Delta \omega\Delta T\ge\frac{1}{2}$, with $\Delta T$ smaller than the Hubble time. The precise form of spectral densities, however, is determined by de Sitter geometry.
Note that at the level of individual frequency modes, $\omega$'s are conserved in scattering processes. The respective
$\delta(\omega_1{-}\omega_2{-}\omega_3)$ in Eq.(\ref{amp1}) reflects the time translational symmetry of embedding space.

The frequency spread narrows when the particles probe very short  (as compared to the curvature scale) invariant spacetime intervals. In the limits described in the previous Section, the densities peak at  $\omega=\sqrt{n^2+\mu^2}$, which in QFT corresponds to the energy $E=\omega$ of a relativistic particle with mass $|\mu|$, which indeed, when $\mu\gg 1$, is the mass of a particle in the principal series representation. Asymptotically,
\be \rho(\omega)~\to~ \frac{\delta(\omega-E)}{\sqrt{2E}\sqrt{2\pi}}\ ,\qquad~ E=\sqrt{n^2+m^2}\ .\ee
In this limit, the scattering amplitudes (\ref{amp1}) become
\be ^{out}\big\langle n_2,\mu_2;n_3,\mu_3\big|n_1,\mu_1\big\rangle^{in}=-i\lambda(2\pi)^2\, \frac{\delta(n_1-n_2-n_3)\, \delta(E_1-E_2-E_3)}{ \sqrt{2E_1}\sqrt{2\pi}\sqrt{2E_2}\sqrt{2\pi} \sqrt{2E_3}\sqrt{2\pi} }\ .\label{amp2}
\ee
Note that the first delta function is Kronecker's because the momenta are discrete while the second one is Dirac's.

In two-dimensional Minkowski spacetime, the three-scalar amplitude in a theory with the scalar potential (\ref{pot3}) is given by \cite{col}:
\be ^{out}\big\langle p_2,m_2;p_3,m_3\big|p_1,m_1\big\rangle^{in}=-i\lambda(2\pi)^2\, \frac{\delta(p_1-p_2-p_3)\, \delta(E_1-E_2-E_3)}{\sqrt{ \langle p_1|p_1\rangle }\sqrt{ \langle p_2|p_2\rangle }\sqrt{ \langle p_3|p_3\rangle }}\ .\label{amin}
\ee
Here, $E$ is the energy and $p$ is the spatial momentum. In two dimensions,  the states are normalized with the inner products
\be\langle p|q\rangle =4\pi E\,\delta(p-q)\ ,\ee
therefore the norm is
\be\sqrt{\langle p|p\rangle} =\sqrt{4\pi E\,\delta(0)}=\sqrt{4\pi E\ell}=\sqrt{2E}\sqrt{2\pi}\ .     \ee
After identifying the momentum $p$ with $n$, we find a perfect agreement between Eqs.(\ref{amin}) and (\ref{amp2}).
The ``high-energy'' amplitudes describing particles probing de Sitter spacetime at very short distances do indeed agree with Minkowski amplitudes. Only one property of spectral densities was used for this conclusion: the delta function-like peak at asymptotically large masses and/or momenta. On physical grounds, we expect this property to hold in any number of dimensions, therefore in this limit, de Sitter amplitudes agree with Minkowski amplitudes. 

\section{Summary and comments on literature}
Given a huge body of work on quantum field theory in de Sitter spacetime, it is important to place our results in the context of previous developments. Most of previous work has been focused on constructing the S-matrix on one of the ``patches:'' either Poincar\'e or static. The restriction to a fixed patch is not only a matter of using specific coordinate systems but most importantly, it changes the global properties of spacetime. It violates global de Sitter invariance. Without maximal symmetry, there is no connection between distinct observers (at least between some of them), and the computation of scattering amplitudes must be supplemented by specifying who is the one measuring a given amplitude.
We circumvented this problem by working in global de Sitter spacetime which is endowed with a priviledged class of ``inertial'' geodesic observers, all related by de Sitter symmetry transformations. Their time evolution is generated by the boost operators of the embedding Minkowski spacetime.

With the boosts generating proper time translation along observers' timelike geodesics, we defined the Hamiltonian of quantum field theory as the Noether charge of the conserved current obtained by contracting the stress-energy tensor with the corresponding Killing vector (\ref{tkil}). In this way, we constructed the quantum evolution operator (\ref{tev}). We were able to cast the  S-matrix elements (in the basis of de Sitter symmetry group representations) in a Dyson's form (\ref{dys}) which is particularly convenient for computing the amplitudes in perturbation theory. To the best of our knowledge, this is the first derivation of de Sitter S-matrix from first principles.

Recently, the authors of  Refs.\cite{Melville:2023kgd,Melville:2024ove,Donath:2024utn} developed a formalism for constructing the scattering amplitudes within the framework of cosmological correlators \cite{Chen:2009zp,Arkani-Hamed:2015bza} on the Poincar\'e patch. The Poincar\'e patch covers the region of $X_0>-X_1$, therefore one expects  agreement between the wavefunctions of quantum states in the upper half of the hyperboloid $(X_0>0)$. Indeed, in the limit of large $n$, when the wavefunctions probe short distances, Eq.(\ref{hank}) agrees with Refs.\cite{Melville:2023kgd,Melville:2024ove,Donath:2024utn} after identifying the momentum $ |\vec{p\,}|\equiv |n|$. Due to different global structures, however, this is all what can be compared -- in addition to the flat limit, in which both formalisms should reproduce Minkowski S-matrix elements. 

We should also mention at least one previous attempt at constructing S-matrix in global de Sitter space \cite{Marolf:2012kh}. Since it utilizes the LSZ formalism, it is technically very different, but it would be  interesting to compare both approaches.

\section{Conclusions}
We live in curved spacetime. The reason why QFT formulated in flat Minkowski spacetime works so well is that the laboratory and accelerator experiments probe spacetime intervals (distances and times)  much shorter than the curvature scale.  Under those circumstances, spacetime appears to be flat, but only in the first approximation. A better approximation is by a spacetime with positive constant curvature. The general case of constant curvature is difficult to study in the absence of some guiding symmetry principles similar to Poincar\'e symmetry of Minkowski spacetime. de Sitter spacetime, however, is more tangible, because its maximal symmetry provides a powerful tool for formulating QFT, by using it in a  similar way as Poincar\'e symmetry is being used in the flat case.
In this work, we developed a formalism for computing the scattering amplitudes in de Sitter spacetime. It can be used for studying the effects of curvature in processes occuring at microscopic as well as at macroscopic scales.

Only very light particles, with masses not too far beyond the cosmological scale of $10^{-33}$ eV ({\it i.e}.\ very low Compton frequencies), are affected by spacetime curvature, and only in processes, in which they propagate with very low momenta ({\it i.e}.\ very long wavelenths). Their wavepackets have a large frequency spread.
In nature, only photons and perhaps neutrinos (and hypothetical gravitons) have masses below or comparable to the curvature. Physics of ultrasoft photons is entangled with spacetime geometry, but it is not clear if the frequency fluctuations discussed in this work are observable. 
Assuming that at least one neutrino is ``cosmologically'' light, another interesting application of our results is to neutrino physics, in particular to neutrino oscillations. We leave it to future work.

The processes involving soft particles play very important role in celestial holography \cite{Strominger:2017zoo,Pasterski:2021rjz,Raclariu:2021zjz}. In this framework, they are described by ``soft theorems'' which are related to the symmetries of asymptotically flat spacetime. Our work shows that in curved spacetime, these theorems no longer apply, and they need to be replaced by some new principles. By reformulating soft theorems, we may learn how to extend celestial holography to curved spacetime.

Our main result, the Dyson's formula (\ref{dys}) for scattering amplitudes in de Sit{\nolinebreak}ter spacetime, holds in arbitrary number of dimensions. Likewise, the general framework for constructing particle wavefunctions by using group-theoretical methods applies to any number of dimensions. Some details, however, were worked out here only in the context of deS$_2$, which served as our  ``proof of concept.'' We leave to the sequel \cite{tbz} a detailed discussion of  deS$_4$. It is  necessary for incoporating, at the quantitative level,  the curvature effects in the standard model of particle physics. 
\section*{Acknowledgements}
TRT is deeply indebted to Laurent Freidel for long discussions during the 2024 annual meeting of the Simons Collaboration on Celestial Holography in New York. Without his advice and encouragement, this work would have never been completed. He is also grateful to Jan Derezi\'nski and Jerzy Lewandowski for enlightening conversations. Unfortunately, Jurek passed away before this work was completed. He will be greatly missed.
This work was supported in part by NSF PHY-2209903, the Simons Collaboration on Celestial Holography, and
Polish National Agency for Academic Exchange under the NAWA Chair programme, and by the Royal Society.
It was also supported by the MAESTRO grant no.\ 2024/54/A/ST2/00009
funded by the National Science Centre, Poland.
Any opinions, findings, and conclusions or
recommendations expressed in this material are those of the authors and do not necessarily
reflect the views of the National Science Foundation.
\appendix
\section{Associated Legendre functions}
The {\bf associated Legendre functions} are solutions of the differential equation
\begin{equation}
(1-z^2) \frac{d^2 u}{d z^2} -2 z \frac{du}{d z} +\left[ \nu(\nu+1) - \frac{\mu^2}{1-z^2}\right] u = 0 \, ,
\end{equation}
where the parameters $\mu$ and $\nu$ are referred to as the order and degree, respectively. In general, they are complex numbers. 
They are defined on the complex plane with a branch cut running from  from $-{}\infty$ to $+{}1$.
There are two kinds of linearly independent associated Legendre functions: $P$ and $Q$.  They are usually defined through  the hypergeometric functions $_2F_1$:
\begin{equation}
P^{\mu}_{\nu}(z) = \frac{1}{\Gamma(1-\mu)} \left( \frac{z+1}{z-1}\right)^{\mu/2} \, _2F_1\left( -\nu, \nu+1; 1-\mu; \frac{1-z}{2} \right) \, ,
\end{equation}
\begin{equation}
Q^{\mu}_{\nu}(z) = \frac{e^{\mu\pi i}\Gamma(\nu+\mu+1)\Gamma(\frac{1}{2})}{2^{\nu+1}\Gamma\left(\nu+\frac{3}{2}\right)} (z^2-1)^{\mu/2} z^{-\nu-\mu-1} \,_2F_1\left(\frac{\nu+\mu+2}{2}, \frac{\nu+\mu+1}{2}; \nu+\frac{3}{2}; \frac{1}{z^2} \right) \, .
\end{equation}
$P$ and $Q$ are also called associated Legendre functions of the first and second kind, respectively.

{\bf Ferrer's functions} are defined on the interval $[-1,1]$ so that $z=x$ ($x=\sin t$ in our case), with the linear combinations of the associated Legendre functions taken above and below the cut:
\begin{align}
P^{\mu}_{\nu}(x) &= \frac{1}{2}\left[ e^{\frac{1}{2} \mu\pi i} P^{\mu}_{\nu}(x+i0) + e^{-\frac{1}{2} \mu\pi i} P^{\mu}_{\nu}(x-i 0)\right] \nonumber\\
&= \frac{1}{\Gamma(1-\mu)} \left( \frac{1+x}{1-x}\right)^{\mu/2} \, _2F_1\left( -\nu,\nu+1; 1-\mu; \frac{1-x}{2} \right) \, ,
\end{align}
\begin{align}
Q^{\mu}_{\nu}(x) & = \frac{1}{2} e^{-\mu\pi i}\left[e^{-\frac{1}{2} \mu\pi i} Q^{\mu}_{\nu}(x+i0) +e^{\frac{1}{2} \mu\pi i} Q^{\mu}_{\nu}(x-i0) \right] \\
&=\frac{\pi}{2\sin \mu \pi} \left[P^{\mu}_{\nu}(x) \cos \mu \pi - \frac{\Gamma( \nu+\mu+1)}{\Gamma(\nu-\mu+1)} P^{-\mu}_{\nu}(x) \right] \, .
\end{align}
The following two identies were often used in the main text:
\begin{equation}
P^{\mu}_{-\nu-1}(z) = P^{\mu}_{\nu}(z) \, ,
\end{equation}
\begin{equation}
Q^{\mu}_{-\nu-1}(x) = \frac{\sin[ (\nu+\mu)\pi]}{\sin [ (\nu-\mu)\pi]} Q^{\mu}_{\nu}(x) - \frac{\pi \cos\nu\pi \cos\mu \pi}{\sin[ (\nu-\mu)\pi]} P^{\mu}_{\nu}(x) \, .
\end{equation}
For the large degree asymptotics, $\nu\gg1$, we used
\begin{equation}
\nu^{-\mu} P^{\mu}_{\nu}(\cos\varphi) = \sqrt{\frac{2}{\nu\pi \sin\varphi}} \cos\left[ \left( \nu+\frac{1}{2}\right) \varphi -\frac{\pi}{4} +\frac{\mu\pi}{2}\right] +O\left(\frac{1}{\sqrt{\nu^3}} \right) \, ,
\end{equation}

\begin{equation}
\nu^{-\mu} Q^{\mu}_{\nu}(\cos\varphi) = \sqrt{\frac{2}{\nu\pi \sin\varphi}} \cos\left[ \left( \nu+\frac{1}{2}\right) \varphi +\frac{\pi}{4} +\frac{\mu\pi}{2}\right] +O\left(\frac{1}{\sqrt{\nu^3}} \right) \, ,
\end{equation}

The {\bf conical functions } also known as {\bf Mehler's functions} are special cases of the associated Legendre when the degree $\mu = -1/2+i\tau$ where $\tau$ is a real number.
In the main text, the following identity was used:
\begin{equation}
P^{-\mu}_{\nu}(z) = \frac{\Gamma(\nu-\mu+1)}{\Gamma(\nu+\mu+1)}\left[P^{\mu}_{\nu}(z) -\frac{2}{\pi} e^{-\mu\pi i}\sin\mu\pi\, Q^{\mu}_{\nu}(z) \right] \, .
\end{equation}

\end{document}